# Design of a Specialized Low Noise Amplifier for Enhancing Non-Classicality in Quantum Applications

Ahmad Salmanogli

Ankara Yildirim Beyazit University, Engineering Faculty, Electrical and Electronic Department, Ankara, Turkey, Ahmadsalmanogli@aybu.edu.tr

**Abstract:**
In this study, we present the design and analysis of a Low Noise Amplifier tailored specifically for quantum applications. We selected the HEMT for its unique noise reduction properties, crucial for quantum engineering. The main goal is to minimize the noise figure within the C-band frequency range (4-8 GHz) to induce nonclassicality in quantum signals. While achieving a noise figure of less than 0.065 dB within this band, we recognized the trade-off with gain, mitigated by incorporating additional stages to maintain noise figure at optimal levels. Quantum analysis of the circuit, employing a simplified model of HEMT due to its complexity, revealed insights into its nonlinear properties and interactions between circuit components and environmental factors. Leveraging Qutip toolbox in Python, we conducted time-evolution analysis of the system, revealing the circuit's behavior as an open quantum system under cryogenic conditions. Our investigation extends to quantifying quantum correlation (quantum discord) and its relationship with noise figure, posing important questions regarding the direct impact of its minimization on circuit nonclassicality at cryogenic temperatures. This comprehensive study sheds light on the intricate interplay between circuit design, and its influence on the relationship between the noise figure and quantum correlation.

**Introduction:**
In quantum applications, the Low Noise Amplifier (LNA) [1-2] plays a critical role in enhancing the performance and reliability of quantum systems. As a fundamental component in signal processing chains, LNAs are tasked with amplifying weak quantum signals while minimizing additional noise, thereby preserving the delicate quantum information [3-10]. By meticulously designing LNAs tailored specifically for quantum environments, such as utilizing High Electron Mobility Transistors (HEMTs) known for their noise reduction properties, researchers aim to achieve optimal signal-to-noise ratios critical for quantum communication, sensing, and computation tasks. Additionally, LNAs engineered to operate under cryogenic conditions further ensure the preservation of quantum coherence and fidelity, enabling the exploration of novel quantum phenomena and facilitating the development of advanced quantum technologies. Thus, the role of LNAs in quantum applications [3-10] extends beyond mere signal amplification, serving as enablers for unlocking the full potential of quantum information processing systems.

The most important element in a LNA, maybe, is the transistor by which the signal amplification is carried out. It may CMOS, HEMT, or other types. HEMTs offer unique advantages that make them particularly well-suited for quantum applications. Unlike conventional CMOS technology [11-12], HEMTs exhibit exceptionally low noise figures (NF), a critical attribute for preserving the delicate quantum information encoded within signals. This property arises from the inherent characteristics of HEMTs, such as their high electron mobility and reduced electron scattering, which contribute to minimal noise generation during signal amplification. Moreover, HEMTs operate efficiently at cryogenic temperatures, aligning with the environmental requirements of many quantum systems. The ability to maintain low noise levels while operating at ultralow temperatures is paramount for sustaining quantum coherence and fidelity, essential for tasks like quantum communication, sensing, and computing. Additionally, the unique nonlinear properties of HEMTs can be harnessed to manipulate quantum states and explore novel quantum phenomena, paving the way for advancements in quantum information processing and technology. Thus, the distinctive features of HEMTs make them indispensable components in the development of cutting-edge quantum devices and systems. In this study, we try to use quantum theory to understand the effect of the HEMT in LAN completely.

Quantum theory [13-19] serves as a powerful analytical tool for dissecting the intricacies of complex systems such as LNAs. By leveraging the principles of quantum mechanics, researchers can gain deeper insights into the behavior of LNAs at the quantum level [13-14]. Quantum analysis provides a framework to understand the subtle interactions between individual components, such as the active elements' nonlinear properties and how they couple with one another and the surrounding environment. This approach allows for a comprehensive examination of the LNA's dynamics, including its quantum correlations and nonclassical behaviors, which are crucial for quantum applications. Moreover, quantum theory enables researchers to predict and optimize the performance of LNAs under various conditions, guiding the design and engineering of next-generation quantum devices with enhanced functionality and efficiency. Thus, quantum theory emerges as an indispensable tool for unraveling the mysteries of LNAs and advancing the frontier of quantum technology. To get to know fully about the quantum correlation, we try to focus on one of the most important quantifiers in quantum realm, which is quantum discord.

Quantum discord [13-14, 20-21] is a measure of the quantum correlations present in a composite quantum system. Unlike classical correlations, which can be fully captured by mutual information, quantum discord accounts for non-classical correlations that cannot be explained by shared classical information alone. Technically, quantum discord quantifies the discrepancy between two different ways of measuring the mutual information: one based on classical correlations and another based on quantum correlations. It reflects the degree to which measurements on one part of the system influence the state of another part in a non-classical manner. In essence, quantum discord captures the intricacies of quantum entanglement and other forms of non-classical correlations, providing valuable insights into the quantum behavior of composite systems. It plays a crucial role in characterizing the quantumness of correlations and is widely used in various quantum information processing tasks and quantum technologies.

This study aims to design an LNA using CAD programs, with a strong focus on minimizing its NF in the C-band. Following this, the study switches to quantum theory to analyze the same circuits fully. Given the complexity of a full analysis, a simplified model of a HEMT and the circuit related is utilized for illustration. Using quantum theory, we calculate various quantum phenomena, such as quantum correlations generated in the circuit. The primary goal is to identify any relationships between NF minimization and the generation of quantum correlations or the enhancement of nonclassicality, based on results from both CAD software and quantum theory analysis.

## System definition and theoretical backgrounds:

*System definition (general view):*

In this study, we have designed a LNA tailored specifically for quantum applications. The architecture comprises a cascode amplifier, integrating Common Source (CS) and Common Gate (CG) stages connected in series. Our choice of HEMT is deliberate, driven by its suitability for quantum applications. Through an extensive literature survey, we have established that HEMT's properties enable significant reduction in NF—a unique advantage that sets it apart from CMOS, particularly in specialized domains like quantum engineering [1, 2].

The schematic and layout modeling were meticulously executed using CAD software. Our primary objective is to minimize the NF within the C-band frequency range (4-8 GHz), with the explicit aim of inducing nonclassicality in quantum signals. Achieving this entails aligning the NF with the noise level of Josephson Parametric Amplifiers (JPAs) under ideal conditions. By doing so, we markedly enhance the probability of nonclassicality emergence at the ultralow temperature of 4.2 K—an accomplishment of profound significance for quantum applications, where quantum correlation serves as a critical parameter for enhancing device performance. The results of the simulation of the layout in ADS show that the NF in the C-band is less than 0.065 dB, which is very suitable for quantum applications, but it is not comparable with the JPA's NF, which is on the order of 0.006 dB. In addition, we lost the gain to survive the NF; however, by considering the main goal it is not an important factor. Indeed, using another stage coupled to the cascade stage, the gain of the circuit can be enhanced, while the NF is maintained at the lowest level. In addition to the simulation of the circuit using CAD software, the circuit is analyzed using quantum theory. Using this method adds some degrees of freedom to completely analyze the circuit, by which one can get to know about the individual part of the circuit and how the active element's nonlinear properties (HEMT) can affect the dynamics equation of motions, how the stages can couple to each other and finally importantly how they couple to the environment created due to the mismatching. In the CAD simulations (schematic and layout depicted in Fig. 1), the nonlinear model of the HEMT is utilized. However, due to the intricate nature of the model, a simplified version is employed for quantum analysis, as illustrated in Fig. 2. This simplified circuit comprises three LC oscillators interconnected through the HEMT nonlinear circuit: RF_input+$LC_1$+HEMT+$LC_2$+HEMT+$LC_3$. The presence of the LC resonators is attributed to bias and matching network effects. The small signal model of the HEMT facilitates coupling between the oscillators. Leveraging quantum theory, we analyze this model, deriving its total classical Hamiltonian. Subsequently, employing the Legendre transformation, we express the total quantum Hamiltonian in terms of ladder operators.

In this study, we utilize Qutip in Python [24] to examine the time evolution of the system, which is defined as an open quantum system. Furthermore, we theoretically derive the quantum Langevin equation to effectively analyze the open quantum system. Leveraging this, we construct the covariance matrix and employ the input-output formula to calculate the circuit's gain. Within this defined system, the LNA design interacts with its environment, operating under cryogenic conditions. An attempt is made to compute the quantum correlation (quantum discord) for the first and third oscillators. This analysis aims to investigate how the circuit design, cascade topology, and the inherent HEMT nonlinearity influence the generated quantum correlation.

Consequently, we utilize the results from both CAD simulations and quantum analysis to address a fundamental question: What is the relationship between the NF and quantum correlation (quantum discord)? Does minimizing the NF directly impact the nonclassicality in the circuit operating at cryogenic temperatures?

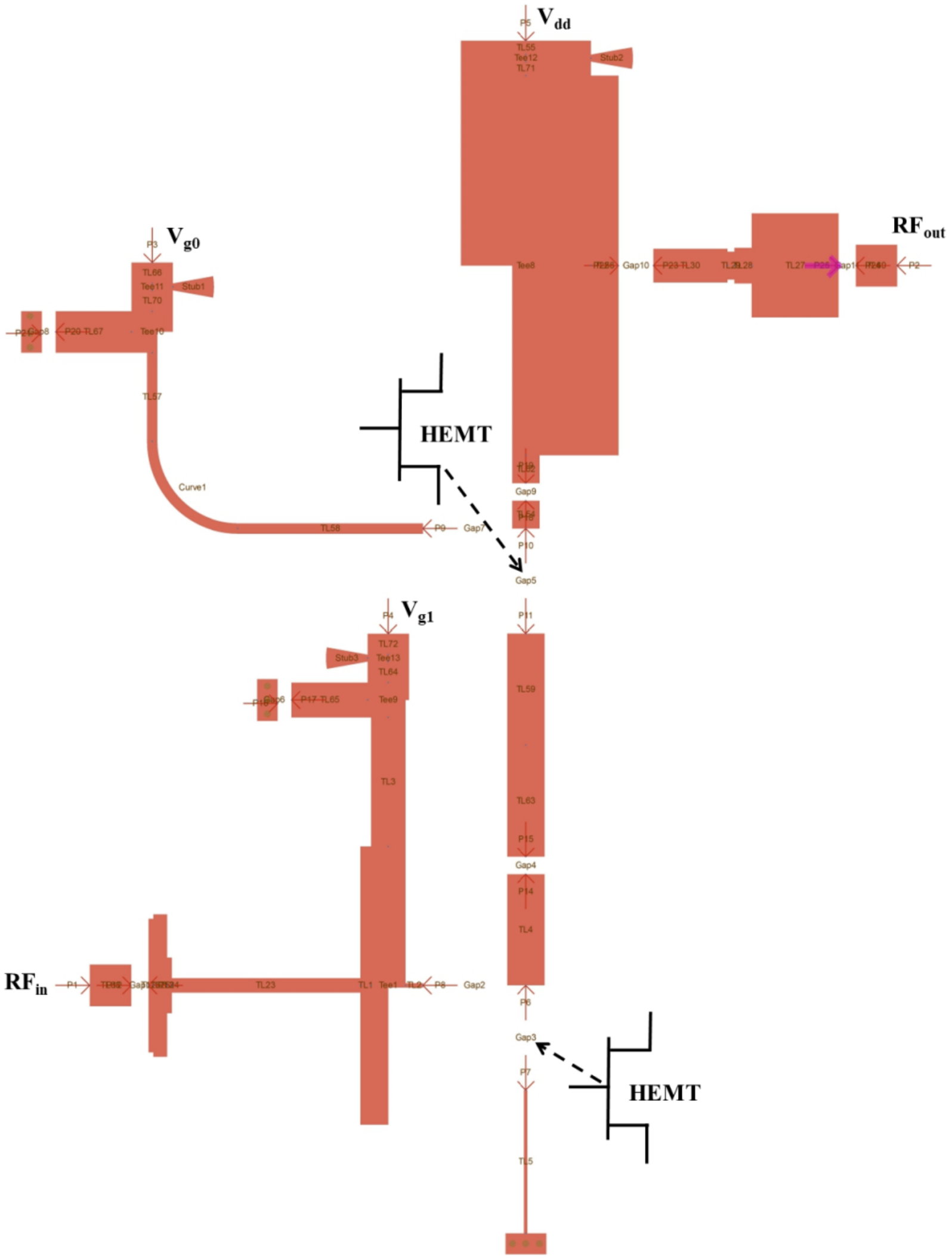

Fig. 1 the layout of the LNA designed using the distributed elements for matching and bias circuits and also a few lumped elements for input and output coupling and stability aims.

*System analysis using Quantum theory*

In this section, we aim to analyze the circuit depicted as a LNA in Fig. 1 with complete accuracy using quantum theory [25, 26]. The circuit includes some bias networks, input and output matching networks, and two nonlinear elements as HEMTs. It is designed to minimize the use of lumped elements, favoring the use of transmission lines (TLs) as distributed elements. However, TLs can be modeled using lumped elements; using transmission line theory, a TL can be modeled with N elements ($L = N*l_e$, where $l_e$ is the length of each element and L is the total length of the TL). Each element consists of a series resistor and inductor in parallel with a capacitor and conductor. If we assume that $l_e$ is less than the excitation wavelength ($\lambda$) and the element is lossless, we can ignore the resistor and conductors from the model. This method simplifies a complex circuit, allowing us to apply quantum theory to fully understand the interactions between the stages, the circuit's operation as an amplifier, and the correlation of modes between different stages. Additionally, we explore any potential relationship between NF and quantum correlation. The simplified version of the LNA is shown in Fig. 2, in which three LC resonators are coupled to each other through the nonlinear element used to model the transistor.

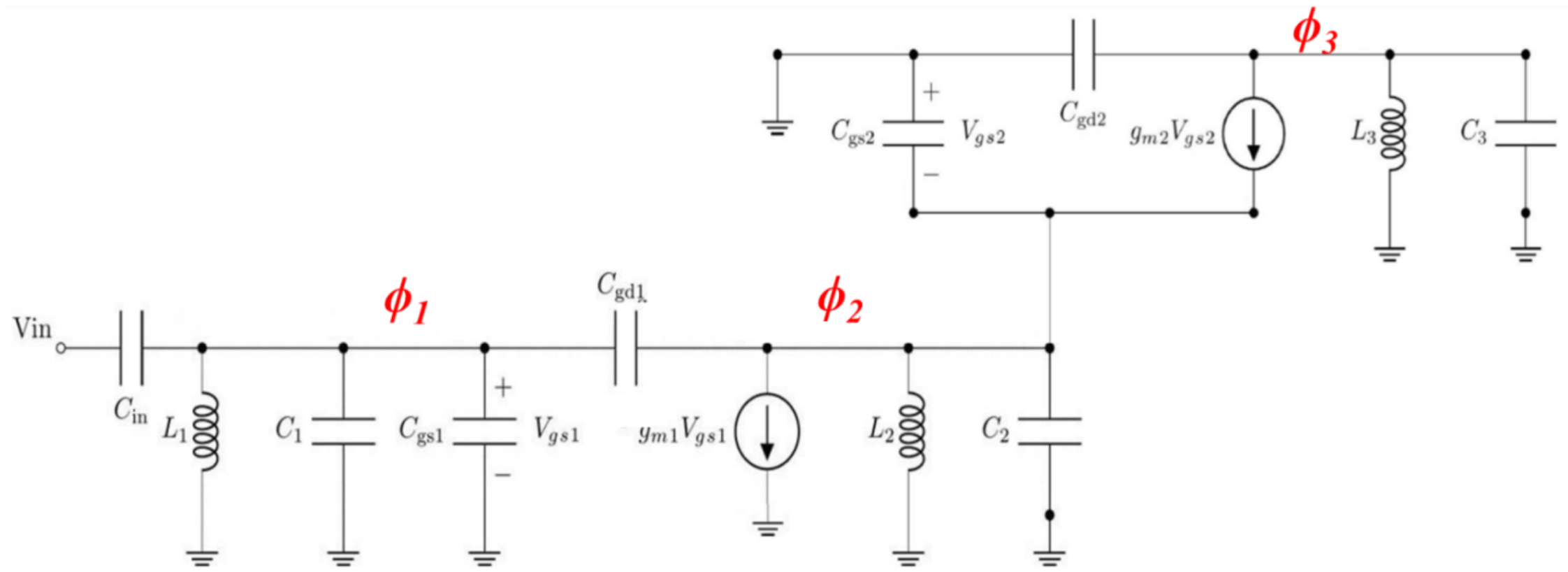

Fig. 2 Small signal model of the circuit; $L_1$, $C_1$, $L_2$, $C_2$, $L_3$, and $C_3$ are generated due to the effect of the bias and matching network in the circuit

The total Lagrangian of the circuit shown is given by:

$$L_t = \frac{C_{in}}{2}\left(V_{in} - \dot{\phi}_1\right)^2 - \frac{1}{2L_1}\phi_1^2 + \frac{C_1 + C_{gs1}}{2}\dot{\phi}_1^2 + \frac{C_{gd1}}{2}\left(\dot{\phi}_1 - \dot{\phi}_2\right)^2 - g_{m1}\dot{\phi}_1\phi_2 - \frac{1}{2L_2}\phi_2^2 + \frac{C_2}{2}\dot{\phi}_2^2$$

$$+\frac{C_{gs2}}{2}\left(-\dot{\phi}_2\right)^2 + g_{m2}\dot{\phi}_2\left(\phi_3 - \phi_2\right) + \frac{C_{gd2}}{2}\left(-\dot{\phi}_3\right)^2 - \frac{1}{2L_3}\phi_3^2 + \frac{C_3}{2}\dot{\phi}_3^2 \tag{1}$$

where $\phi$ is the node flux. To attain the total classic Hamiltonian, one needs to examine the quantum conjugate momentum relating to the each node flux using the Legendre equation [14-19]. Therefore, the quantum Hamiltonian of the circuit analyzed is given by:

$$H_t = \left\{\hbar\omega_1 a_1^+ a_1 + \hbar\omega_2 a_2^+ a_2 + \hbar\omega_3 a_3^+ a_3\right\}$$
$$-\left\{\frac{1}{2C_{Q12}}\frac{\hbar}{2\sqrt{Z_1Z_2}}\left(a_1 - a_1^+\right)\left(a_2 - a_2^+\right) + \frac{1}{2C_{Q13}}\frac{\hbar}{2\sqrt{Z_1Z_3}}\left(a_1 - a_1^+\right)\left(a_3 - a_3^+\right) + \frac{1}{2C_{Q23}}\frac{\hbar}{2\sqrt{Z_2Z_3}}\left(a_2 - a_2^+\right)\left(a_3 - a_3^+\right)\right\}$$
$$-i\left\{\begin{array}{l} G_{12}\frac{\hbar}{2}\sqrt{\frac{Z_2}{Z_1}}\left(a_1 - a_1^+\right)\left(a_2 + a_2^+\right) + G_{13}\frac{\hbar}{2}\sqrt{\frac{Z_3}{Z_1}}\left(a_1 - a_1^+\right)\left(a_3 + a_3^+\right) + G_{23}\frac{\hbar}{2}\sqrt{\frac{Z_2}{Z_3}}\left(a_2 + a_2^+\right)\left(a_3 - a_3^+\right) \\ +G_{32}\frac{\hbar}{2}\sqrt{\frac{Z_3}{Z_2}}\left(a_2 - a_2^+\right)\left(a_3 + a_3^+\right) + G_{33}\frac{\hbar}{2}\sqrt{\frac{Z_3}{Z_3}}\left(a_3 - a_3^+\right)\left(a_3 + a_3^+\right) + \frac{i}{2L_{23p}}\sqrt{Z_2Z_3}\frac{\hbar}{2}\left(a_2 + a_2^+\right)\left(a_3 + a_3^+\right) \end{array}\right\} \tag{2}$$
$$+V_{in}C_{in}\left\{C_{11}\left(-i\sqrt{\frac{\hbar}{2Z_1}}\left(a_1 - a_1^+\right)\right) + 2C_{12}\left(-i\sqrt{\frac{\hbar}{2Z_2}}\left(a_2 - a_2^+\right)\right) + 2C_{13}\left(-i\sqrt{\frac{\hbar}{2Z_3}}\left(a_3 - a_3^+\right)\right) - C_{12}g_{m2}\sqrt{\frac{\hbar Z_3}{2}}\left(a_3 + a_3^+\right)\ldots\right\}$$

where $\omega_1$, $\omega_2$, $\omega_3$, $Z_1$, $Z_2$, and $Z_3$ are respectively, the angular frequency and effective impedance of the first, second, and third oscillators, respectively; Also $C_{Q12}$, $C_{Q13}$, $C_{Q23}$, $G_{12}$, $G_{13}$, $G_{23}$, $G_{32}$, $G_{33}$, $L_{23p}$, $C_{11}$, $C_{12}$, and $C_{13}$ are the variable parameters summarized in Appendix 1. In the following analysis, we use the derived Hamiltonian to evaluate both the quantum characteristics of the system and the gain produced. For gain calculation, we derive the quantum Langevin equation and transform the equations into the Fourier domain. By constructing the scattering matrix, the gain of the signal and the idler can be determined [23]. It is important to note that the idler is generated due to the nonlinearity of the HEMT in the system. To analyze the quantum features of the system, QuTiP toolbox is utilized [21]. Alongside the total quantum Hamiltonian, we need to define the system-environment interaction Hamiltonian and the environment Hamiltonian. Additionally, it is necessary to describe the system operators ($a_1$, $a_2$, and $a_3$), the environment operators, and the thermal bath operator to fully analyze the Lindblad master equation using QuTiP [24].

Solving the Lindblad master equation provides the sub-elements of the covariance matrix, such as $\langle a_1^+ a_1\rangle$, $\langle a_1 a_2^+\rangle$, $\langle a_1 a_3\rangle$, and so on. This allows us to establish the covariance matrix for each pair of oscillators, such as ($LC_1$ and $LC_2$) and ($LC_1$ and $LC_3$), and calculate the associated eigenvalues and eigenvectors. These quantities are crucial for thoroughly analyzing the features of a quantum system. For instance, the covariance matrix can be used to calculate and analyze the quantum correlations between the oscillators, providing deeper insights into the system's quantum behavior and performance. This comprehensive approach is essential for understanding and optimizing quantum systems for advanced applications. For instance the covariance matrix calculated can be used to calculate and analyze the quantum correlation between the oscillators. Following, it is tried to illustrate the results derive using the quantum theory and classical analysis using a CAD.

**Results and Discussions:**

This section aims to present the results from both quantum theory simulations and CAD simulations, and to identify a common ground between them. The dc characterization of the HEMT is depicted in Fig. 3a, with the DC point indicated by a black dashed arrow. This characterization helps determine the operating point of the transistor, essential for optimizing its performance in the circuit. It provides valuable insights into the transistor's behavior under different biasing conditions, aiding in efficient circuit design and analysis. One of the important parameters related is the NF and minimum NF relevant to the design shown in Fig. 3b. The NF in the band discussed is found to be less than 0.07 dB, indicating excellent suitability for quantum applications. Low NF is crucial for preserving the integrity of quantum signals, ensuring minimal signal degradation and maximizing signal-to-noise ratio for accurate measurements. A key outcome of this work is our intuitive prediction that with such a low noise figure, the likelihood of achieving non-classicality at the output or quantum correlation between the oscillators is increased.

Fig. 3c displays the input and output reflection coefficients, showcasing their values across the frequency spectrum. In the range of 4-5.5 GHz, the reflection coefficients are below -5 dB, indicating a fair impedance matching. Beyond this range, they remain below -10 dB, suggesting effective suppression of reflections due to the matching network in the circuit. It is clear that proper impedance matching minimizes signal loss and maximizes power transfer efficiency in the circuit. The stability factor of the circuit, as the other important parameter, is illustrated in Fig. 3d, indicating values greater than 1. A stability factor exceeding unity (>1) indicates unconditional stability, signifying that the circuit remains stable under all operating conditions. This ensures reliable and robust performance, free from oscillations or instability issues that could compromise circuit functionality. Fig. 3e presents the fundamental and third harmonic output powers, highlighting the significant difference between them, exceeding 80 dB. This large separation between the fundamental and third harmonic signals indicates effective harmonic suppression, crucial for maintaining signal purity and minimizing spectral contamination. It ensures that the amplifier primarily amplifies the desired signal while suppressing unwanted harmonic distortions. Finally Fig. 3f depicts the spectrum of the designed amplifier, showcasing peaks at specific frequencies such as 5 GHz and 10 GHz, along with their corresponding amplitudes. These peaks represent the amplified signal frequencies and their harmonics. Understanding the spectral characteristics of the amplifier aids in frequency planning and interference mitigation, ensuring optimal performance in practical applications.

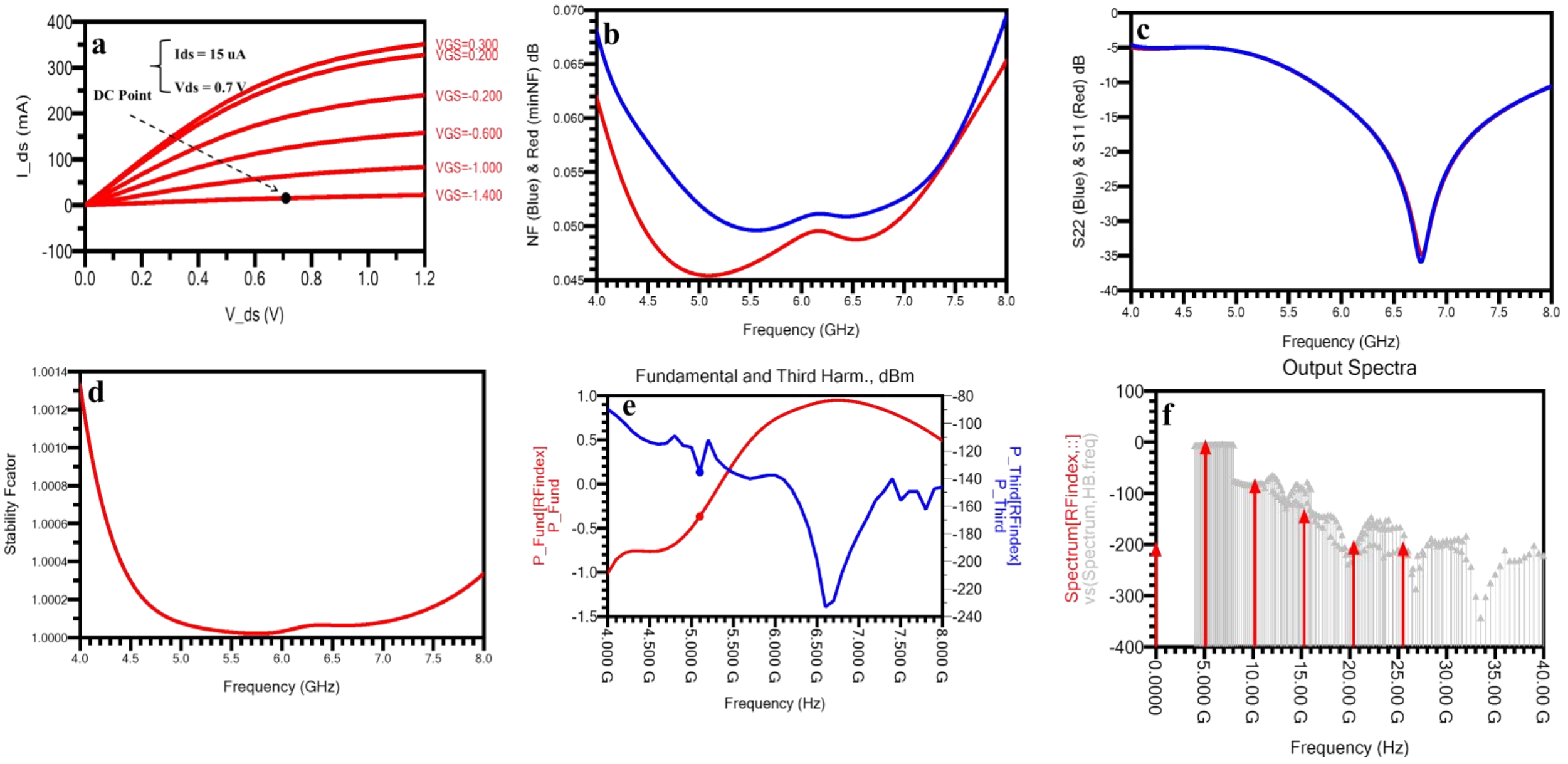

Fig. 3a) dc characterization of the HEMT used in the circuit; DC point selected is indicated by black dashed arrow on the figure; b) NF and minimum NF related to the design in the C-band; C) the input and output reflection coefficients; d) the circuit stability factor; e) the fundamental and third harmonic output powers; f) the spectrum of the amplifier showing some related peaks in 5 GHz, 10 GHz.

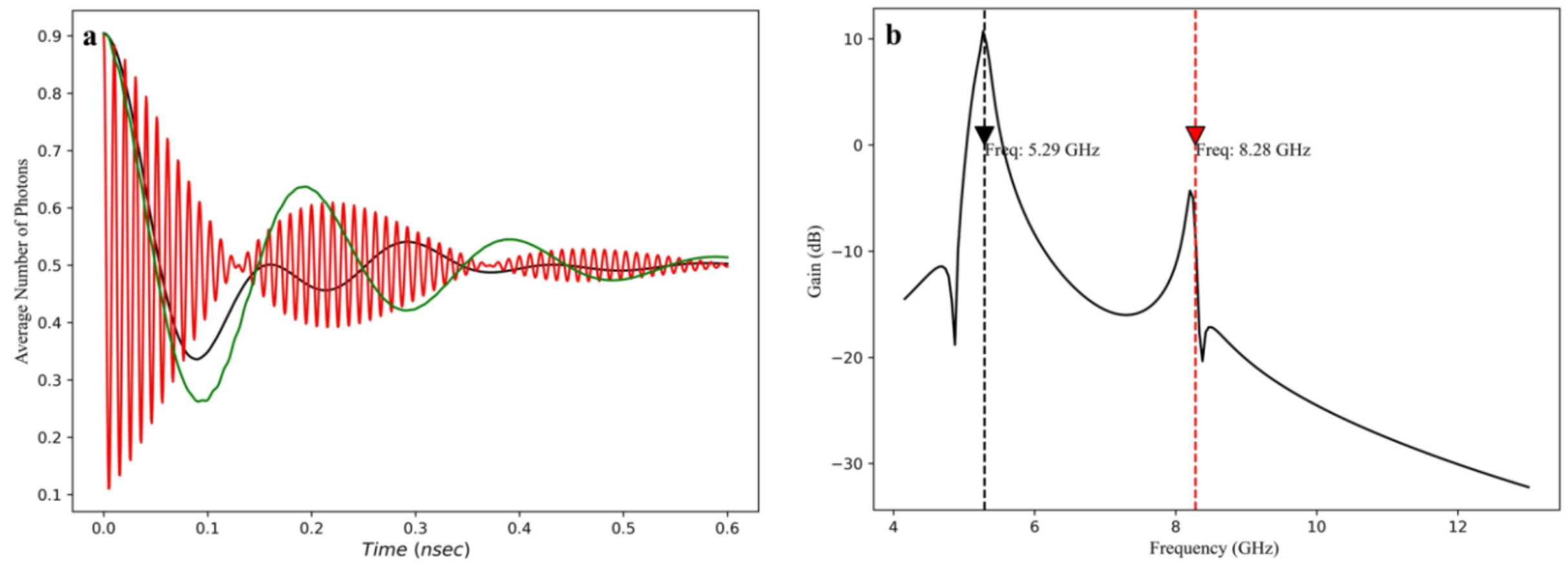

Fig. 4a) the time evolution of the average number of photons (green: $LC_1$, black: $LC_3$, and red: $LC_2$); b). Gain (dB) vs. Frequency (GHz) shows two distinct peaks, while the dominant one is located so close to 5.2 GHz.

After presenting the results of the CAD simulations, particularly the NF of the designed circuit, the outcomes of the quantum theory analysis are then discussed. The time evolution of the average number of photons, as depicted in Fig. 4a, was simulated by solving the Lindblad Master equation using the Qutip toolbox. The plot shows the average photon numbers for three successive resonators: $LC_1$ (green), $LC_3$ (black), and $LC_2$ (red). First of all, the decaying behavior observed indicates the interaction of the resonators with the surrounding environment. However, the most notable behavior arises from $LC_2$, which exhibits a mixing behavior by inherently coupling $LC_1$ and $LC_3$ modes to each other. This coupling effect is crucial in understanding the dynamics of the circuit, its behavior under different conditions, and also the nonlinearity induced by HEMT. In Fig. 4b, the gain of the circuit analyzed using quantum theory is presented. Two distinct peaks are observed, with the dominant peak closely aligned with the exciting frequency of 5.2 GHz. The discrepancy between the quantum theory and CAD simulation results, where the gain in the C-band approaches unity, can be attributed to the simplified linear model of the HEMT used in the quantum analysis. For the sake of simplicity, we had to use a simple model of the HEMT in quantum analysis. Regarding the second peak at 8.2 GHz in the gain graph, it likely emerges due to the mixing behavior observed in $LC_2$. This mixing behavior can introduce additional resonance frequencies and complex interactions between the resonators, leading to the appearance of secondary peaks in the gain spectrum. The presence of this second peak suggests the presence of higher-order resonant modes or nonlinear effects within the circuit, indicating a more intricate behavior than what is captured by the linear model used in the quantum analysis.

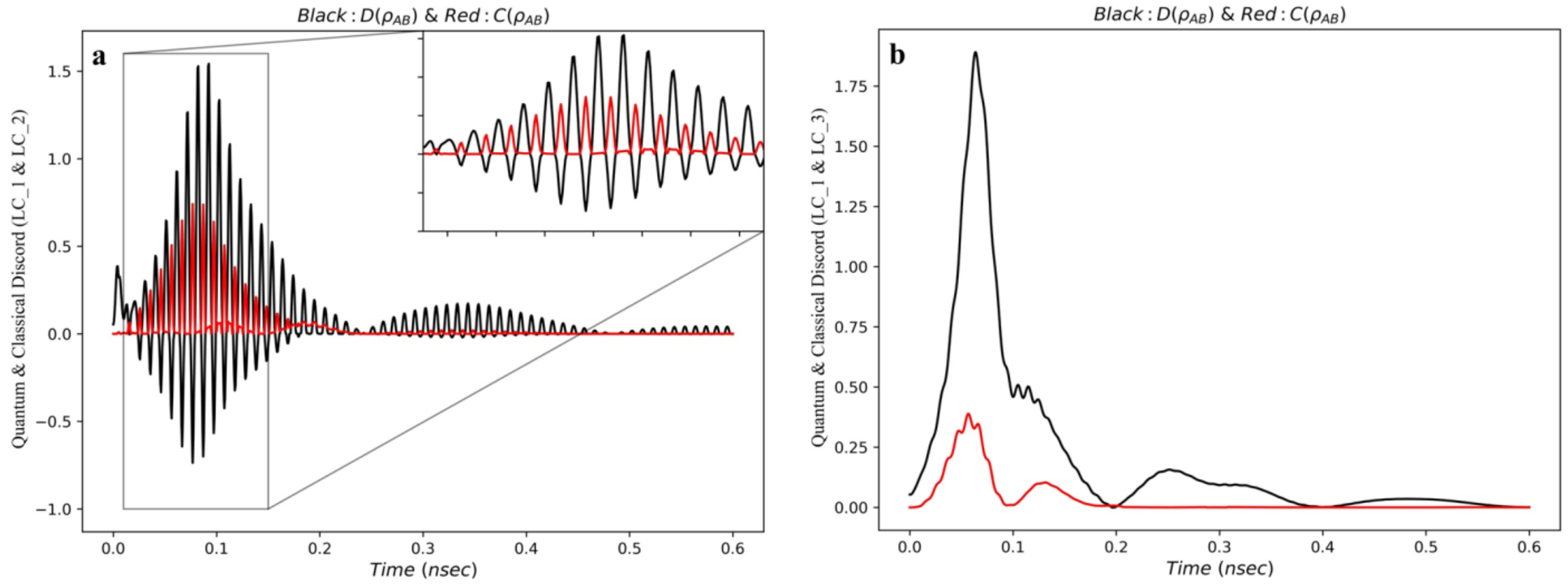

Fig. 5a) the time evolution of the quantum and classical discord between a) the first & second oscillators; b) the first & third oscillators

In addition to calculating the average photon number and circuit gain, another crucial aspect relevant to quantum applications such as quantum discord is computed, and is depicted in Fig. 5. The figures compare both quantum and classical discords with each other. The presence of quantum discord indicates the existence of quantum correlations that cannot be explained by classical physics, while classical discord represents classical correlations that are independent of quantum effects. Results indicate that for time intervals less than 0.2 nanoseconds, the quantum discord surpasses unity (>1), suggesting quantum correlation between modes [20-22], such as $LC_1$ and $LC_2$ in Fig. 5a, and $LC_1$ and $LC_3$ in Fig. 5b, ultimately enhancing nonclassicality at the output. This nonclassicality behavior arises from the circuit's nonlinearity effect. Nonclassicality refers to phenomena in quantum realm that cannot be explained by classical physics. Quantum correlations, such as quantum entanglement or quantum discord, are key indicators of nonclassical behavior. When quantum discord is present between $LC_1$ and $LC_3$, it suggests that these modes are quantum correlated in a way that cannot be explained classically. This heightened quantum correlation enhances the nonclassicality of the output signals, indicating that the system is exhibiting behavior beyond classical physics. Therefore, by generating quantum correlation between $LC_1$ and $LC_3$, the nonclassicality of the output is indeed increased, reflecting the system's quantum nature. Notably, the inset in Fig. 5a highlights that where quantum discord peaks, classical discord is minimized. This phenomenon occurs because quantum and classical correlations often behave inversely in certain quantum systems. Moreover, the gradual decrease in signal amplitude is attributed to device-medium interactions. From a classical sense, the coupling to environment is arisen due to any mismatching between environments. The simulation underscores the circuit's potential shown in Fig. 1 to generate quantum correlations among microwave photons, facilitated by minimized NF in the band operation. Consequently, such circuits can hold promise for inducing nonclassicality in quantum applications at cryogenic temperatures.

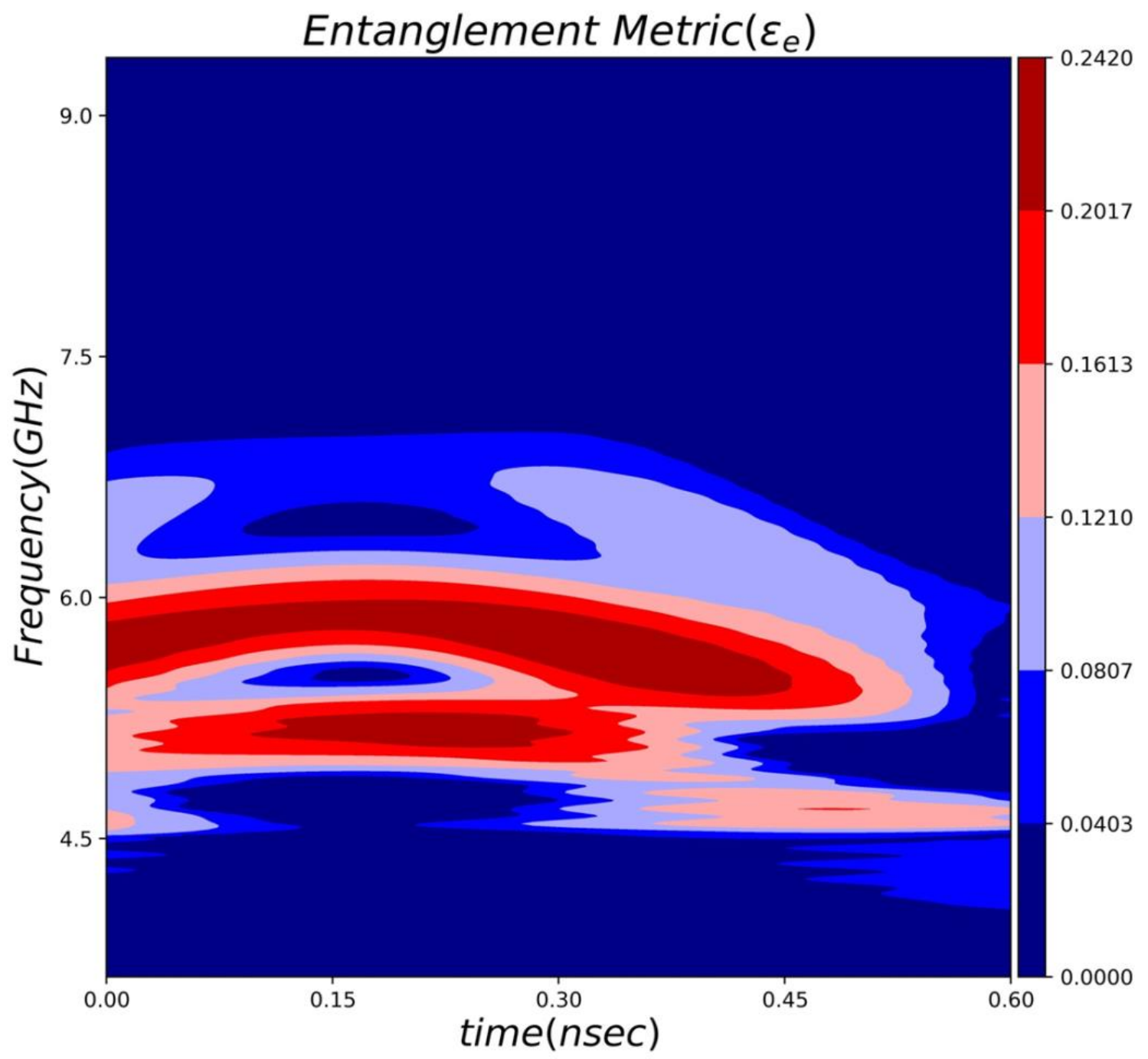

Fig. 6 Entanglement vs frequency (GHz) and time (nsec)

However, to address the main question of this article which is "does minimizing the NF directly impact the nonclassicality in the circuit operating at cryogenic temperatures?" The results regarding the entanglement (entanglement metric $\varepsilon_e$ as a quantifier [25, 26]) between $LC_1$ and $LC_3$ depicted in Fig. 6 can present logic an answer. On can compare Fig. 6 and its peaks with the gain graph in Fig. 4b, which reveals two distinct peaks, one of them dominant around 5.29 GHz, in the frequency range of 4.7 to 5.8 GHz. This observation suggests a correlation between gain enhancement and the emergence of entanglement between the two resonators. The presence of these peaks can be attributed to the design strategy aimed at minimizing the NF while maximizing gain, which in turn amplifies the quantum correlation between the signals. Specifically, around the frequency of 5.2 GHz, there exists a notable potential to augment the quantum correlation. However, it is necessary to note that achieving entanglement between $LC_1$ and $LC_3$ represents a significant milestone in quantum circuit design. This entanglement signifies the establishment of nonclassical correlations in the circuit. Of course, it should be noted that the amplitude of $\varepsilon_e$, as an entanglement metric, is so less than the unity. This means that a strong entanglement between modes mentioned didn't be established it is maybe due to the loss introduced by the circuit, nonlinearity of the HEMT, or extra noise generated because of the incomplete matching networks.

**Conclusions:**

This study presents a comprehensive design and analysis of a LNA tailored for quantum applications, with a focus on minimizing the NF within the C-band frequency range (4-8 GHz). Leveraging the unique noise reduction properties of HEMTs, the circuit achieves an impressive NF of less than 0.065 dB, essential for inducing nonclassicality in quantum signals. The DC characterization of the HEMT and extensive simulations using CAD software and quantum theory reveal the circuit's robust performance, stability, and effective harmonic suppression. In the other hand, the quantum analysis, employing a simplified HEMT model, unveils the intricate interactions within the circuit, demonstrating the dynamic coupling between resonators and the environment. Using QuTiP toolbox, we evaluated the time evolution of the average photon number and gain, highlighting the circuit's potential for maintaining low NF while addressing gain trade-offs. The investigation into quantum correlations, particularly quantum discord, underscores the circuit's capability to enhance nonclassicality by fostering quantum correlations between modes. The results indicate a significant relationship between NF and quantum correlation, with minimized NF directly impacting the emergence of nonclassical behavior at cryogenic temperatures. The study also emphasizes the role of the circuit's design in achieving entanglement, albeit limited by circuit losses and extra noise. These findings provide valuable insights into optimizing quantum circuits for advanced applications, showcasing the critical interplay between circuit design, NF, and quantum correlation in the pursuit of high-performance quantum systems. This work sets the stage for further exploration and refinement in the field of quantum engineering.

**Conflict of interest:** There is no conflict of interest for this work.

**Ethics approval and consent to participate:** I confirm that this work is original and has been neither published nor is currently under consideration for publication elsewhere.

**Consent for publication:** The author of this study gives the publisher the permission of the author to publish the work.

**Availability of data and materials:** There are no datasets generated for this work; however, the code generated during the current study is available from the corresponding author upon reasonable request.

**Competing interests:** There are no competing interests.

**Funding:** There is no funding for this work.

**Authors' contributions:** All of the studies have been done by Ahmad Salmanogli.

**Appendix 1:**

In this appendix all of the variables used in Eq. 2 listed as follows: $C_{Q12}$, $C_{Q13}$, $C_{Q23}$, $G_{12}$, $G_{13}$, $G_{23}$, $G_{32}$, $G_{33}$, $L_{23p}$, $C_{11}$, $C_{12}$, and $C_{13}$ are introduced:

$$\begin{bmatrix} \dot{\phi}_1 \\ \dot{\phi}_2 \\ \dot{\phi}_3 \end{bmatrix} = \begin{bmatrix} C_{11} & C_{12} & C_{13} \\ C_{21} & C_{22} & C_{23} \\ C_{31} & C_{32} & C_{33} \end{bmatrix} \begin{bmatrix} Q_1 \\ Q_2 \\ Q_3 \end{bmatrix} - \begin{bmatrix} 0 & C_{11}g_{m1} - C_{12}g_{m2} & C_{12}g_{m2} \\ 0 & C_{21}g_{m1} - C_{22}g_{m2} & C_{22}g_{m2} \\ 0 & C_{31}g_{m1} - C_{32}g_{m2} & C_{33}g_{m2} \end{bmatrix} \begin{bmatrix} \phi_1 \\ \phi_2 \\ \phi_3 \end{bmatrix} - \begin{bmatrix} -C_{11}C_{in} & 0 & 0 \\ -C_{21}C_{in} & 0 & 0 \\ -C_{31}C_{in} & 0 & 0 \end{bmatrix} \begin{bmatrix} V_{in} \\ 0 \\ 0 \end{bmatrix}$$

$$C_{N1} = C_1 + C_{in} + C_{gs1} + C_{ds1}$$

$$C_{N2} = C_2 + C_{ds1} - C_{gs2}$$

$$C_{N3} = C_3 + C_{ds2} + C_{out}$$

$$\frac{1}{C_{q1}} = \frac{C_{N1}C_{11}^2 + C_{21}^2 C_{N2} + C_{31}^2 C_{N3} - 0.5C_{11}C_{12}C_{ds1}}{2}$$

$$\frac{1}{C_{q2}} = \frac{C_{N1}C_{12}^2 + C_{22}^2 C_{N2} + C_{32}^2 C_{N3} - 0.5C_{22}C_{12}C_{ds1}}{2}$$

$$\frac{1}{C_{q3}} = \frac{C_{N1}C_{13}^2 + C_{23}^2 C_{N2} + C_{33}^2 C_{N3} - 0.5C_{33}C_{13}C_{ds1}}{2}$$

$$\frac{1}{C_{q1q2}} = C_{N1}C_{11}C_{12} + C_{N2}C_{21}C_{22} + C_{N3}C_{31}C_{32} - C_{ds1}\left(C_{11}C_{22} + C_{12}C_{21}\right)$$

$$\frac{1}{C_{q1q3}} = C_{N1}C_{11}C_{13} + C_{N2}C_{21}C_{23} + C_{N3}C_{31}C_{33} - C_{ds1}\left(C_{11}C_{23} + C_{13}C_{21}\right)$$

$$\frac{1}{C_{q2q3}} = C_{N1}C_{12}C_{13} + C_{N2}C_{22}C_{23} + C_{N3}C_{22}C_{32} - C_{ds1}\left(C_{12}C_{23} + C_{13}C_{22}\right)$$

$$\frac{1}{L_{\phi 2}} = \frac{C_{N1}\left(C_{11}g_{m1} - C_{12}g_{m2}\right)^2 - C_{N2}\left(C_{21}g_{m1} - C_{22}g_{m2}\right)^2 - C_{N3}\left(C_{31}g_{m1} - C_{32}g_{m2}\right)^2 - C_{ds1}\left(C_{11}g_{m1} - C_{12}g_{m2}\right)\left(C_{21}g_{m1} - C_{22}g_{m2}\right)}{2}$$

$$\frac{1}{L_{\phi 3}} = \frac{C_{N1}\left(C_{12}g_{m1}\right)^2 + C_{N2}\left(C_{22}g_{m2}\right)^2 + C_{N3}\left(C_{32}g_{m2}\right)^2 - C_{ds1}\left(C_{21}C_{22}g_{m2}^2\right)}{2}$$

$$g_{12} = C_{N1}C_{11}\left(C_{11}g_{m1} - C_{12}g_{m2}\right) + C_{N2}C_{21}\left(C_{21}g_{m1} - C_{22}g_{m2}\right) + C_{N3}C_{31}\left(C_{31}g_{m1} - C_{32}g_{m2}\right) - C_{ds1}\left\{\begin{aligned} &C_{11}\left(C_{21}g_{m1} - C_{22}g_{m2}\right) \\ &+ C_{21}\left(C_{11}g_{m1} - C_{12}g_{m2}\right) \end{aligned}\right\}$$

$$g_{22} = C_{N1}C_{12}\left(C_{11}g_{m1} - C_{12}g_{m2}\right) + C_{N2}C_{22}\left(C_{21}g_{m1} - C_{22}g_{m2}\right) + C_{N3}C_{32}\left(C_{31}g_{m1} - C_{32}g_{m2}\right) - C_{ds1}\left\{\begin{aligned} &C_{12}\left(C_{21}g_{m1} - C_{22}g_{m2}\right) \\ &+ C_{22}\left(C_{11}g_{m1} - C_{12}g_{m2}\right) \end{aligned}\right\}$$

$$g_{32} = C_{N1}C_{13}\left(C_{11}g_{m1} - C_{12}g_{m2}\right) + C_{N2}C_{23}\left(C_{21}g_{m1} - C_{22}g_{m2}\right) + C_{N3}C_{33}\left(C_{31}g_{m1} - C_{32}g_{m2}\right) - C_{ds1}\left\{\begin{aligned} &C_{13}\left(C_{21}g_{m1} - C_{22}g_{m2}\right) \\ &+ C_{23}\left(C_{11}g_{m1} - C_{12}g_{m2}\right) \end{aligned}\right\}$$

$$g_{13} = C_{N1}C_{11}C_{12}g_{m2} + C_{N2}C_{21}C_{22}g_{m2} + C_{N3}C_{31}C_{32}g_{m2} - C_{ds1}\left\{C_{11}C_{22}g_{m2} + C_{12}C_{21}g_{m2}\right\}$$

$$g_{23} = C_{N1}C_{11}C_{12}g_{m2} + C_{N2}C_{21}C_{22}g_{m2} + C_{N3}C_{32}C_{32}g_{m2} - C_{ds1}\left\{C_{12}C_{22}g_{m2} + C_{12}C_{22}g_{m2}\right\}$$

$$g_{33} = C_{N1}C_{13}C_{12}g_{m2} + C_{N2}C_{23}C_{22}g_{m2} + C_{N3}C_{33}C_{32}g_{m2} - C_{ds1}\left\{C_{13}C_{22}g_{m2} + C_{12}C_{23}g_{m2}\right\}$$

$$\frac{1}{2C_{Q1}} = C_{11} - \frac{1}{C_{q1}}, \frac{1}{2C_{Q2}} = C_{22} - \frac{1}{C_{q2}}, \frac{1}{2C_{Q3}} = C_{33} - \frac{1}{C_{q3}},$$

$$\frac{1}{C_{Q12}} = C_{21} + C_{12} - \frac{1}{C_{q1q2}}, \frac{1}{C_{Q13}} = C_{31} + C_{13} - \frac{1}{C_{q1q3}}, \frac{1}{C_{Q23}} = C_{32} + C_{23} - \frac{1}{C_{q2q3}},$$

$$G_{12} = \left(-g_{m1}C_{11} - g_{m2}C_{12}\right) + g_{12}, G_{13} = -g_{m2}C_{12} + g_{13}, G_{23} = -g_{m1}C_{13} - \left(g_{m1}C_{31} - g_{m2}C_{32}\right) + g_{32},$$

$$G_{22} = -g_{m1}C_{12} - \left(g_{m1}C_{21} - g_{m2}C_{22}\right) + g_{22}, G_{33} = -g_{m2}C_{32} + g_{33},$$

$$\frac{1}{2L_2} = -g_{m1}\left(C_{11}g_{m1} - C_{12}g_{m2}\right) - \frac{1}{L_{\phi 2}}, \frac{1}{2L_{3p}} = \frac{1}{2L_3} - \frac{1}{L_{\phi 3}}$$

$$Z_{11} = \sqrt{\frac{L_{11}}{C_{Q1}}}, Z_{22} = \sqrt{\frac{L_2}{C_{Q2}}}, Z_{33} = \sqrt{\frac{L_{3p}}{C_{Q3}}}, \omega_{11} = \frac{1}{\sqrt{L_{11}C_{Q1}}}, \omega_{22} = \sqrt{\frac{1}{L_2C_{Q2}}}, \omega_{33} = \sqrt{\frac{1}{L_{3p}C_{Q3}}}$$